# A Motion Assessment Method for Reference Stack Selection in Fetal Brain MRI Reconstruction Based on Tensor Rank Approximation

Haoan Xu[1], Wen Shi[1,2], Jiwei Sun[1], Tianshu Zheng[1], Cong Sun[3], Sun Yi[4], Guangbin Wang[5], and Dan Wu[1]

[1] Key Laboratory for Biomedical Engineering of Ministry of Education, Department of Biomedical Engineering, College of Biomedical Engineering & Instrument Science, Zhejiang University, Hangzhou, China

[2] Department of Biomedical Engineering, Johns Hopkins University School of Medicine, Baltimore, MD, United States

[3] Department of Radiology, Beijing Hospital, National Center of Gerontology, Institute of Geriatric Medicine, Chinese Academy of Medical Sciences, P.R. China.

[4] MR Collaboration, Siemens Healthcare China, Shanghai, China

[5] Department of Radiology, Shandong Provincial Hospital Affiliated to Shandong First Medical University, Jinan, Shandong, China

Correspondence to:

Dan Wu, Ph.D.

Room 525, Zhou Yiqing Building, Yuquan Campus,

Zhejiang University, Hangzhou, 310027, China

E-mail: danwu.bme@zju.edu.cn

Phone: +86-15825508242

## Abstract

**Purpose:** Slice-to-volume registration and super-resolution reconstruction (SVR-SRR) is commonly used to generate 3D volumes of the fetal brain from 2D stacks of slices acquired in multiple orientations. A critical initial step in this pipeline is to select one stack with the minimum motion as a reference for registration. An accurate and unbiased motion assessment (MA) is thus crucial for successful selection.

**Methods:** We presented a MA method that determines the minimum motion stack based on 3D low-rank approximation using CANDECOMP/PARAFAC (CP) decomposition. Compared to the current 2D singular value decomposition (SVD) based method that requires flattening stacks into matrices to obtain ranks, in which the spatial information is lost, the CP-based method can factorize 3D stack into low-rank and sparse components in a computationally efficient manner. The difference between the original stack and its low-rank approximation was proposed as the motion indicator.

**Results:** Compared to SVD-based methods, our proposed CP-based MA demonstrated higher sensitivity in detecting small motion with a lower baseline bias. Experiments on randomly simulated motion illustrated that the proposed CP method achieved a higher success rate of 95.45% in identifying the minimum motion stack, compared to SVD-based method with a success rate of 58.18%. We further demonstrated that combining CP-based MA with existing SRR-SVR pipeline significantly improved 3D volume reconstruction.

**Conclusion:** The proposed CP-based MA method showed superior performance compared to SVD-based methods with higher sensitivity to motion, success rate, and lower baseline bias, and can be used as a prior step to improve fetal brain reconstruction.

## Key words

motion assessment; fetal brain MRI; singular value decomposition; tensor decomposition; slice-to-volume registration.

# 1. Introduction

Fetal brain MRI is a powerful tool in prenatal diagnosis and developmental neuroscience for its high spatial resolution and strong tissue contrast [1,2], compared to ultrasound [3]. However, due to the irregular fetal movement and maternal abdominal motion, fetal brain MRI is often subject to various motion artefacts [4,5]. While fast 2D multi-slice imaging methods, such as single-shot fast spin-echo (SSFSE) or balanced steady state free precession (bSSFP), are commonly used to freeze intra-slice motion [6], inter-slice motion is still inevitable and remains challenging for retrospective 3D volume reconstruction [7].

The current volume reconstruction pipeline utilizes 2D slices of fetal brain MRI acquired in multiple orientations and performs iterative slice-to-volume registration (SVR) and super-resolution reconstruction (SRR) [8]. In each iteration, the SVR step registers individual slices to a target 3D volume to correct the misalignment introduced by inter-slice motion, and then the SRR step generates isotropic high-resolution volumes from the re-aligned stacks. In this SVR-SRR pipeline, a reference stack with the minimum motion needs to be determined as the template for initial volume-to-volume registration between the stacks, and also as the template for SVR after being interpolated into an isotropic volume [9]. As shown in Supplementary Figure 1, a motion-corrupted reference stack will make it difficult to find motion parameters for some slices, and the incorrect registration and rejection of slices will affect the final reconstructed volume. Previously, the reference stack was determined manually or by simply using the first input stack [10]. Therefore, an effective motion assessment (MA) method is desired for the optimal performance of a fully automated fetal brain processing pipeline.

Several MA methods have been proposed. Jiang et al. proposed a motion detection method for functional MRI based on motion parameters obtained from the registration between successive volumes [11], which was only applicable to functional MRI data and not suitable for our purpose. Atkinson et al. presented the use of an entropy focus criterion to quantify the degree of motion artefact, as motion decreased the number of dark pixels and thus reduced the entropy [12]. This method, nevertheless, was only suitable for detecting large-scale intra-slice motion. Tourbier et al. integrated a motion index computation algorithm based on tracking the centroid displacement of input image masks in their super-resolution toolkit [13], which was only sensitive to translational motion but not rotational motion. Moreover, there had been an increasing interest in applying machine learning techniques in MA. Küstner et al. utilized a convolutional neural network to extract deep features that contain motion information between adjacent slices [14]. Butskova et al. used adversarial Bayesian optimization to infer the distribution of motion parameters and trained a regressor for quantifying the motion artefact within slices [15]. Lorch et al. proposed an automatic detection framework based on the random forest to evaluate synthesized head and respiratory motion artefacts [16]. But it was known that these methods highly relied on a large number of training data and their generalizability was always under concern.

Recently, Kainz et al. proposed a motion estimation method for fetal brain MRI based on singular value decomposition (SVD) that flattened slices into vectors as rows of a data matrix for SVD [17]. The motion indicator (MI) was determined by the rank of data matrix, which was calculated as the error between the original matrix and the approximation matrix reconstructed from the first few singular values and vectors [18]. As a larger stack was supposed to have more anatomical information and thus was more suitable for reference, Ebner et al. modified it by normalizing the MI with respect to the effective volume of different stacks [9] and integrated the modified algorithm into the NiftyMIC toolkit. Applying SVD to a huge matrix (e.g., $\mathbf{D} \in \mathbb{R}^{20 \times 10000}$ for a stack of size $100 \times 100 \times 20$) was time-consuming, so GPU acceleration was used in Kainz's work. Moreover, the decomposition of a 3D stack into a 2D matrix inevitably lost spatial information [19], and the stacks in different orientations might not have similar baseline ranks because of the inherent structural difference. More importantly, for the fetal brain at high gestational age (GA), the structure became more complex, and the flattened matrix was usually full rank, making it difficult to compare the motion between stacks.

In this work, we introduced a tensor decomposition method, named

CANDECOMP/PARAFAC decomposition [20,21], to factorize a 3D stack into low-rank and sparse components to extract its motion information. We also compared it with an improved version of Kainz's method by performing SVD on re-sliced images along all three axes to make for the loss of spatial information due to stack flattening. We hypothesized that the CP-based method would be sensitive to motion by making full use of motion information in 3D, and relatively unbiased to stacks in different orientations at baseline through interpolating stacks into isotropic volumes. To evaluate the performance of different MA methods, we simulated linearly increasing and random motions onto motion-free fetal brain volumes and then tested the correlation between proposed motion indicators and the simulated motions.

## 2. Methods

### 2.1 Theory

We compared the proposed 3D CP decomposition-based MA with the modified 2D SVD based on re-sliced stacks (SVD-RSS), as well as the NifityMIC version of Kainz's method with flattened stack (SVD-FS) in Figure 1. In this section, we first briefly describe the principle of matrix rank and the relationship between low-rankness and inter-slice motion in Section 2.1.1. Then, the principles of SVD and two different implementations are described in Section 2.1.2. Finally, the principle of tensor rank and CP decomposition is described in Section 2.1.3. The code of CP and SVD-RSS methods is available at https://github.com/xuhaoan/FetalMotionAssess. SVD-FS is performed directly using the NiftyMIC toolkit (https://github.com/gift-surg/NiftyMIC).

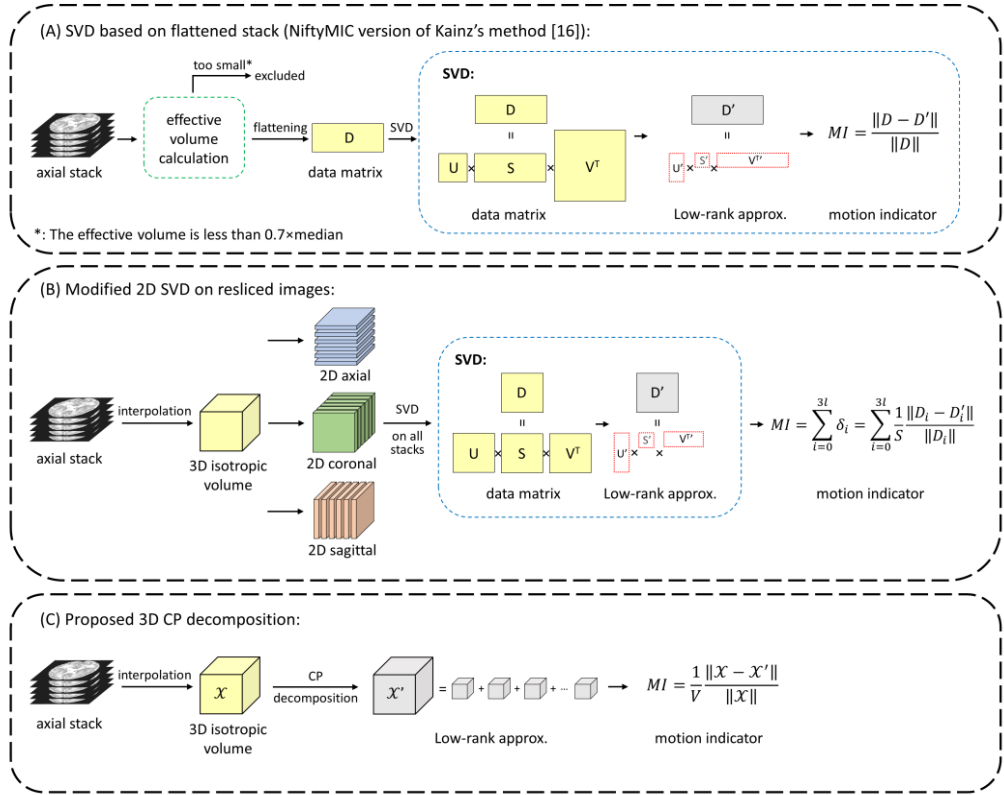

**Figure 1:** Pipelines of three MA methods. (A) NifityMIC version of Kainz's method which flattens 2D slices into 1D vectors as rows of data matrix for decomposition. (B) Proposed re-slicing SVD-based method. The input stack is interpolated into 3D volume and resliced along three orientations for 2D SVD. (C) Proposed CP-based method. The data volume obtained by interpolation is decomposed into *r* rank-one tensors to obtain motion information.

### 2.1.1 Rank and low-rank approximation

The rank of a matrix measures the linear correlation of rows and columns [22]. In MRI data, two adjacent slices share similar structural information. Therefore, for a motion-free stack, the well-aligned slices are linearly correlated, and the 3D volume or 2D flattened matrix of this stack should be low-rank. The rank is expected to increase upon inter-slice motion. Low-rank approximation provides a matrix with a smaller rank compared to the original matrix to obtain a compact representation of the data with limited loss of information [23].

### 2.1.2 SVD based on re-sliced stacks (SVD-RSS)

SVD can provide the low-rank approximation of a 2D matrix [24,25]. SVD factorizes the data matrix $\mathbf{D} \in \mathbb{R}^{m \times n}$ into two orthogonal singular vector matrices $\mathbf{U} \in \mathbb{R}^{m \times m}$, $\mathbf{V}^T \in \mathbb{R}^{n \times n}$ and a rectangular diagonal matrix $\mathbf{S} \in \mathbb{R}^{m \times n}$ whose singular values are in descending order.

$$\mathbf{D}_{m \times n} = \mathbf{U}_{m \times m} \mathbf{S}_{m \times n} \mathbf{V}^T_{n \times n} \quad (1)$$

The number of nonzero singular values in $\mathbf{S}$ determines the rank of data matrix $\mathbf{D}$. To achieve low-rank approximation, the first $r < \min(m, n)$ singular vectors and $r$ singular values are selected to reconstruct a rank-$r$ matrix $\mathbf{D}'$ through

$$\mathbf{D}'_{m \times n} = \mathbf{U}_{m \times r} \mathbf{S}_{r \times r} \mathbf{V}^T_{r \times n} = \sum_{i=0}^{r} s_i \mathbf{u}_i \mathbf{v}_i^T \quad (2)$$

where $s_i$ is a singular value in $\mathbf{S}$, $\mathbf{u}_i$ and $\mathbf{v}_i^T$ are singular vectors in $\mathbf{U}$ and $\mathbf{V}^T$. The approximate matrix $\mathbf{D}'$ represents the low-rank component of the original data matrix, and the error $\mathbf{D} - \mathbf{D}'$ is the sparse component that corresponds to high-frequency noise caused by fetal motion and structural variance.

In our modified version SVD-RSS method, the input stack is first interpolated into an isotropic volume to minimize the bias between the stacks acquired in different orientations. Then, as shown in Figure 1B, we reslice the volume into 2D stacks along x/y/z axes and perform SVD on the three sets of re-sliced 2D images. This step is added given the fact that images along different axes provide complimentary motion information. In the principal axis, as shown in Supplementary Figure S2A, fetal motion leads to translation and rotation between slices; whereas in non-principal axes, as shown in Supplementary Figure S2B-C, fetal motion mainly results in sheared or distorted images, both of which will increase the matrix rank. To balance the accuracy and efficiency, for each SVD, we select five singular values and vectors to reconstruct a rank-5 approximate matrix, and the rank selection will be specifically described in section **2.2.4**.

The relative error between the original matrix and the approximate matrix is used as the indicator of fetal motion. Specifically, we use the sum of differences between approximate and original images as a motion indicator (MI). For an isotropic volume with the size of $l \times l \times l$ interpolated from a stack, the relative errors in all stacks

along the three axes are added up to obtain the MI:

$$\mathbf{MI}_{SVD} = \sum_{i=0}^{3l} \frac{1}{S} \frac{\|\mathbf{D}_i - \mathbf{D}_i'\|}{\|\mathbf{D}_i\|} \qquad (3)$$

where $\mathbf{D}_i$ and $\mathbf{D}_i'$ is the original and approximate data matrix, respectively. $\|\cdot\|$ represents Frobenius norm. The effective area of corresponding slice $S$ is introduced to normalize the through-plane motion, considering through-plane motion is likely to increase the effective area of re-sliced images along non-principal axes thus leading to a decrease of MI.

### 2.1.3 CP-based method

SVD is only applicable for decomposing 2D matrices while a stack is essentially a 3D tensor containing tens of 2D slices, so we introduce a tensor decomposition method, namely CANDECOMP/PARAFAC or CP decomposition, which can be considered as a higher-order principal component analysis method, to handle the 3D tensor [26]. CP decomposition factorizes a tensor into a sum of rank-one tensors, and the rank-one tensor can be formulated as the outer product of three vectors [27]. So, given a tensor $\mathcal{X} \in \mathbb{R}^{k \times m \times n}$ with $k$ slices in it, the approximate tensor can be written as:

$$\mathcal{X}' = [\![\lambda; \mathbf{A}, \mathbf{B}, \mathbf{C}]\!] = \sum_{i=1}^{r} \lambda_i \mathbf{a}_i \circ \mathbf{b}_i \circ \mathbf{c}_i \qquad (4)$$

where $r$ is the number of rank-one tensors. ∘ represents the outer product of the vector. $\lambda_i$, $\mathbf{a}_i$, $\mathbf{b}_i$, $\mathbf{c}_i$ are weight and three factor vectors, respectively. $\mathbf{A}$, $\mathbf{B}$, $\mathbf{C}$ are factor matrices that contain a combination of factor vectors. The alternating least squares (ALS) algorithm is utilized to solve the CP decomposition [28,29].

The rank of a tensor is defined as the minimal number of rank-one tensors whose sum is equivalent to the original tensor [26]. The sum of rank-one tensors can be considered as a low-rank component, and MI is defined as the relative error between the original tensor and the approximate tensor that contains a series of rank-one tensors:

$$\mathbf{MI}_{CP} = \frac{1}{V} \frac{\|\mathcal{X} - \mathcal{X}'\|}{\|\mathcal{X}\|} \qquad (5)$$

where $V$ is the effective volume of data tensor, which is used to normalize the through-plane motion. In the proposed CP-based method, we also interpolate the input stack into an isotropic volume to minimize the influence of different resolutions along three axes before CP decomposition.

### 2.2 Experiment

#### 2.2.1 Data acquisition and preprocessing

A total of 180 fetal brain MRI (GA: 20.4 – 40.0 weeks) with at least three orientations (axial, coronal, and sagittal) were collected on a 3T Siemens Skyra scanner (Siemens Healthineers, Erlangen, Germany) with an abdominal coil. The images were acquired with a T2-weighted half-Fourier single-shot turbo spin-echo (HASTE) with the following protocol: repetition time/echo time = 800/97 ms, in-plane resolution = 1.09 × 1.09 mm, field of few = 256 × 200 mm, thickness = 2 mm, partial Fourier factor = 5/8, echo train length = 102, and GRAPPA factor = 2. The axial, coronal, and sagittal images were repeated 1-6 times per orientation.

31 cases were excluded before data preprocessing due to low signal-to-noise ratio, low image quality, or signal voids. The remaining 149 cases were processed with bias field correction by the N4 algorithm [30], brain masking by manual delineation, and 3D non-local means denoising [31]. The preprocessed 2D stacks were used to reconstruct 3D isotropic volumes at a resolution of 0.8 × 0.8 × 0.8 mm using NiftyMIC toolkit [9]. Altogether, 39 cases failed in the SVR-SRR step due to large motion or poor image quality, and a total of 110 high-resolution 3D volumes (GA: 21.7 – 40.0 weeks) were reconstructed successfully, which then were rigidly registered to the spatiotemporal fetal brain atlas [32] using FLIRT [33] and resized into 192 × 192 × 144 with zeros padded in the surrounding regions. These 110 fetal brain volumes with minimum motion were used as ground truth for the following experiments.

#### 2.2.2 Motion simulation

We parameterized the motion of fetal brain by 3D rigid transformation $\boldsymbol{T}$ with 6 Degrees of Freedom (6-DoF) [34]. The motion was presented in the Euler-Cartesian space with a rotation matrix $\boldsymbol{R}$ based on three rotation parameters $\boldsymbol{\theta} = (\theta_x, \theta_y, \theta_z)^T$ and a translation component $\boldsymbol{d} = (d_x, d_y, d_z)^T$ [35]. The transformation parameter $\boldsymbol{T}$ could be written as a 4 × 4 matrix:

$$T(R, d) = \begin{bmatrix} R & d \\ 0 & 1 \end{bmatrix} \quad (6)$$

Transformation parameters for adjacent slices could be considered as a temporal sequence, and a motion trajectory was generated to represent the continuous motion. Slices were sampled from the corresponding position of the transformed volumes [36]. We generated linearly increasing motion and random motion, as specified in the following two sections.

*1) Linear motion*

We simulated large and small degrees of motions to test the sensitivity of different MA methods. The first group had a large motion with rotation of 5° and translation of 1 mm between adjacent slices, according to the average degree of motion we observed in motion-corrupted stacks [36]. The second group had a smaller motion with rotation of 2° and translation of 0.4 mm between adjacent slices. Supplementary Figure S2 illustrates exemplary cases of large and small motion-corrupted stacks. We further simulated a variety of motion degrees by varying the rotation between adjacent slices from 0 to 5° at an interval of 0.5° and in-plane translation from 0 to 2 mm at an interval of 0.2 mm.

*2) Random motion*

We simulated pseudorandom motion trajectories to represent the irregular and complex fetal brain motion in the real world. The motion trajectories were simulated based on a control-point scheme [37], which generated several control points based on a random walk model [38]. To approximate the real-world situation, interleaved acquisition was simulated by combining two stacks with different motion trajectories in an interleaved order ([first stack: 1, 3, 5, … second stack: 2, 4, 6, …]). For a stack with $N$ slices, control points $\{P_1, ... P_{2N}\}$ were generated based on $P_i = P_{i-1} + \Delta P$, where $P_1$ and $\Delta P$ were drawn from specific uniform distributions. $\Delta P$ between two neighboring control points represented the speed and direction of fetal motion. After the generation of control points, smoothing cubic splines [39] were used to fit a motion trajectory. To match the temporal order of interleaved acquisition, the motion trajectory was then split into first and second halves for the simulation of odd and even slices. In this part, the local variation $\Delta T$ of rotation and translation between two adjacent slices was determined by a uniform distribution $U$ with a maximum of 5° and 1 mm, and the global offset $\bar{T}$ was set to twice of local variation. The maximum absolute value was limited to 25° and 5 mm to avoid unrealistic fetal motion. Figure 2 demonstrates a 34-week fetal brain data with interleaved random motions in three stacks simultaneously.

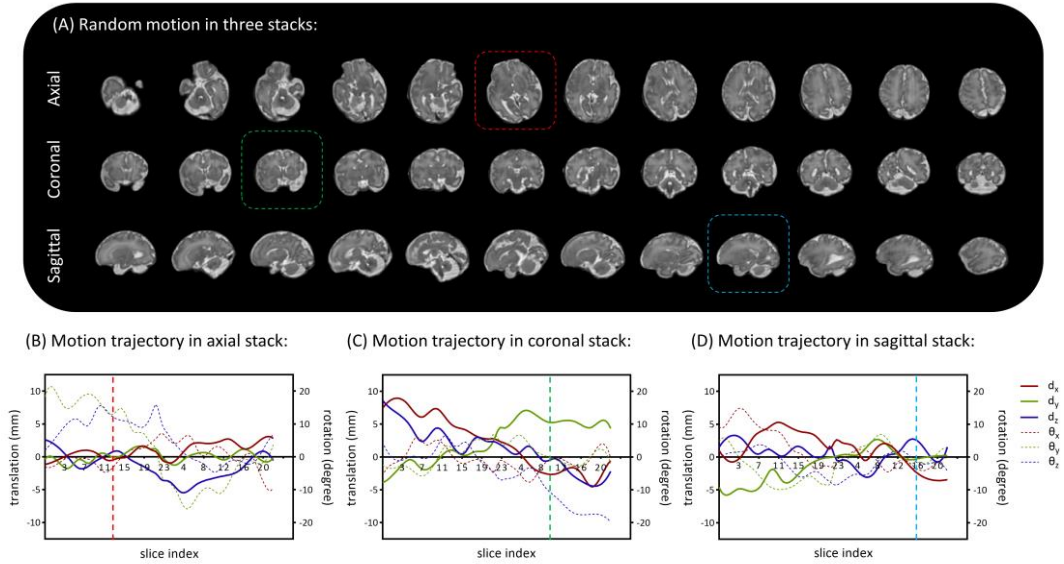

**Figure 2:** Simulation of interleaved random motion in a 34-week-old fetal brain. (A) Motion-corrupted slices in axial (top row), coronal (middle row) and sagittal (bottom row) stacks. (B-D) Corresponding motion trajectories in axial, coronal and sagittal stacks with an interleaved order. Note the slice index is numbered as [1, 3, 5, …, 2, 4, 6, …].

### 2.2.3 Evaluation metrics

To evaluate the performance of different methods, we proposed two metrics, namely relative motion indicator (RMI) and baseline motion indicator (BMI). RMI was the ratio of MIs after and before adding motion and reflects motion detection sensitivity. BMI was the ratio of MIs between different orientations in the absence of added motion, e.g., MI of the coronal and sagittal stacks with respect to the axial stack, which was used to evaluate the degree of bias towards slice orientation.

We also assessed the success rate of selecting the correct reference stack with the minimum motion. For each fetus, we simulated motion trajectories in two stacks, and left one stack with no motion as the motion-free stack.

We further combined the proposed MA step with NiftyMIC pipeline and compared the reconstruction quality to the default pipeline [9] that takes the first input stack (axial stack) as the reference. All reconstructed volumes were rigidly registered to the corresponding ground truth. Structural similarity (SSIM) and normalized root mean square error (NRMSE) between reconstructed volumes and ground truth were used as evaluation metrics to measure the reconstruction quality. Structural dissimilarity (DSSIM) maps were employed to visualize the regional dissimilarity [40].

### 2.2.4 Rank selection

We determined the optimal rank for SVD-RSS and CP according to their performance and computational cost under different ranks. For a given rank, we calculated the average RMI and computational time for 110 fetuses from the large linear motion group. It was worth noting that only in SVD-RSS, the rank was limited to the number of input image rows.

## 3. Result

### 3.1 Optimal rank for CP and SVD-RSS

The result of rank selection experiment showed that for SVD-RSS, RMI first increased with rank and reached its maximum of 1.2481 at rank 5, and then decreased; while the computational time had no significant changes with rank, so we chose 5 as the optimal rank of low-rank approximation. In contrast, RMI of CP showed a monotonic increase with rank, with a faster increase at the beginning and then a slower change towards higher rank, while the computational time almost linearly increased with rank. To balance the accuracy and efficiency, we chose 25 as the number of rank-one tensors, at which the RMI reached 90% maximum performance at rank-50 (rank-25: 1.5390, maximum: 1.5936) and the computational time was about 50% of the maximum.

### 3.2 Performance of MA methods for linear motion

RMI and BMI using the three MA algorithms were compared in Figure 3. In the large-motion group, the proposed CP method outperformed the other two methods with the highest RMI of 1.52, 1.59 and 1.51 for axial, coronal, and sagittal stacks (Figure 3A), as opposed to RMI of 1.24, 1.35 and 1.25 using SVD-RSS and 1.23, 1.20 and 1.49 using SVD-FS. Paired t-test demonstrated significant differences ($P<10^{-4}$) between three methods in all axes. In the small-motion group, CP also achieved significantly higher motion sensitivity ($P<10^{-4}$) and more consistent performance between different orientations than the two SVD-based methods (Figure 3B). Moreover, in Figure 3C, BMI obtained by CP and SVD-RSS in three orientations showed less bias and baseline error between different slice orientations than SVD-FS. The specific values of RMI and BMI were given in the Supplementary Table. S1.

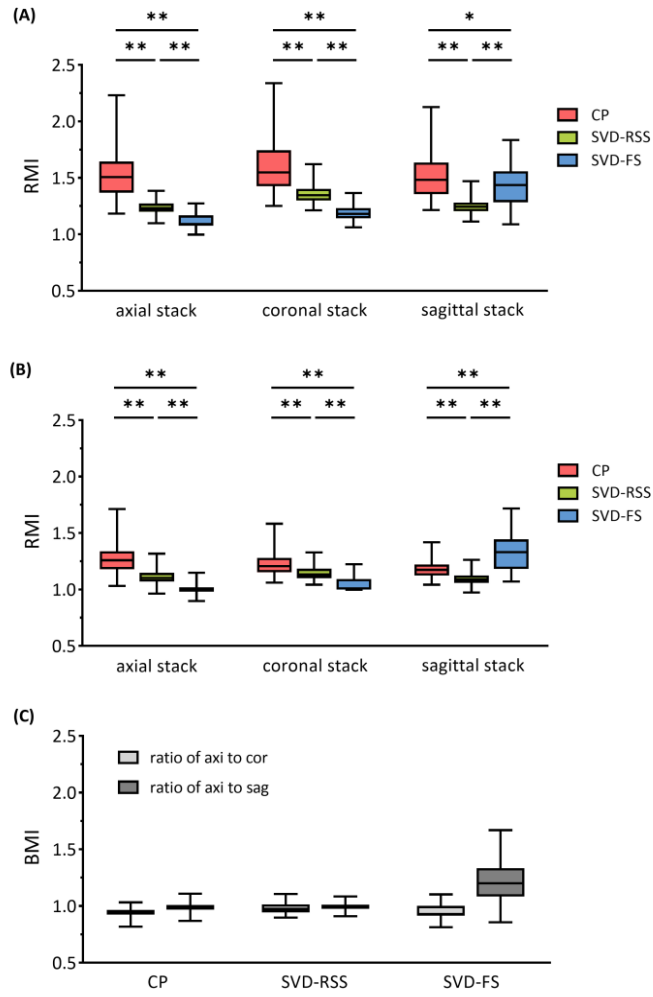

**Figure 3:** Comparison of RMI using CP, SVD-RSS, and SVD-FS methods for detecting large linear motion (A) and small linear motion (B). (C) BMI estimated from motion-free stacks using the three methods. (**$P<10^{-4}$) by paired t-test.

**Table. 1.** Comparison of the success rates using three motion estimation methods in large and small linear motion groups.

| MA method | Large linear motion | | | Small linear motion | | |
| --- | --- | --- | --- | --- | --- | --- |
| | axial | coronal | sagittal | axial | coronal | sagittal |
| CP | **100%** | **100%** | **100%** | **100%** | **96.36%** | **100%** |
| SVD-RSS | **100%** | **100%** | **100%** | 99.09% | 91.82% | 94.55% |
| SVD-FS | 98.18% | 73.64% | 96.36% | 64.55% | 15.45% | 88.18% |

Table. 1 showed a comparison of success rates in large and small motion groups. For large linear motion, both SVD-RSS and CP reached a 100% success rate in determining the minimum motion stacks, while SVD-FS provided success rates of 98%, 74%, and 96% when the minimum motion laid in axial, coronal, and sagittal orientations, respectively. For small motion, the CP method still retained a 100% success rate for axial and sagittal stacks and successfully estimated 96% cases for coronal stacks, outperforming SVD-RSS and SVD-FS methods. The lower success rate in coronal orientation than the other two orientations was likely related to the lower effective volume in coronal stacks that essentially increased the MI and over-estimation of motion.

We further tested RMI for different amounts of translational and rotational motions using three methods, in 110 fetal brains. Figure 4 showed that RMI increased with the increasing translational and rotational motions using all three methods, and CP achieved higher sensitivity compared to SVD-RSS and SVD-FS overall. The RMI curves in Figure 4A-C illustrated almost a linear increase with the translational motion, and CP was the most sensitive to translational motion indicated by the significantly highest slope ($P<10^{-4}$). For rotational motion (Figure 4D-F), the increase of RMI was milder and more complex compared to translation, indicating lower sensitivity for rotation assessment. For translation and rotation along the sagittal axis, SVD-FS showed better performance than SVD-RSS and CP for translation<1 mm and rotation<3°, but CP remained superior for motion above 1 mm or 3°. The reason for this phenomenon would be specifically discussed in Section **4**. 2D heatmaps for simultaneous changes of averaged RMI with translation and rotation along three orientations were shown in Supplementary Figure S3 for three methods.

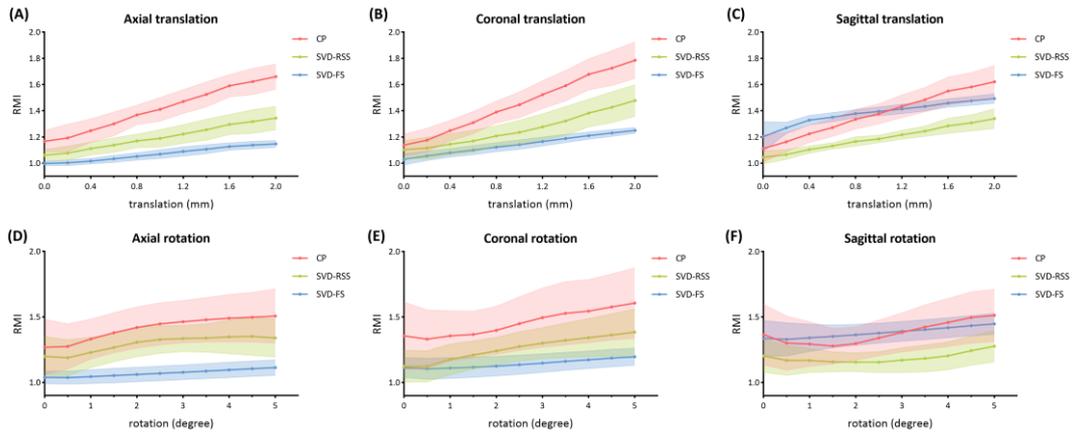

**Figure 4:** RMI curves estimated with varying translational (A-C) and rotational (D-F) motions using CP, SVD-RSS and SVD-FS methods. Shadows indicated the standard deviation from 110 fetal brain data.

Figure 5 showed the SVR-SRR reconstruction of fetal brains in the large and small motion groups. The brains reconstructed by the default pipeline without MA for reference selection were subject to failure or errors, whereas adding the CP-based MA step considerably improved the reconstruction quality (Figure 5A). Paired t-test showed that the volumes reconstructed using CP-selected reference stack had significantly higher SSIM and lower NRMSE with ground truth in both large and small motion groups (Figure 5B-C), compared to that from the default SVR-SRR pipeline.

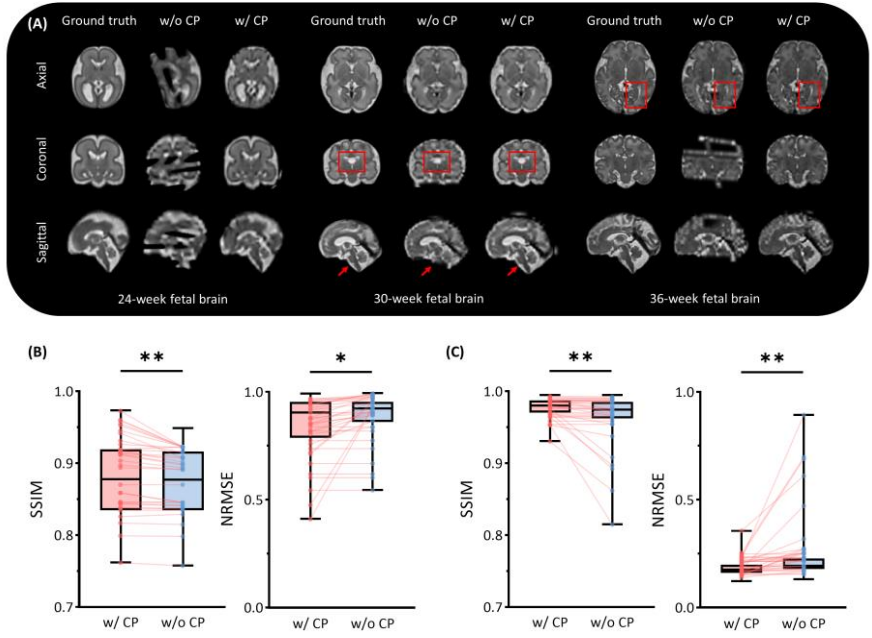

**Figure 5:** (A) Reconstructed 3D fetal brain volumes using CP-based selection of optimal reference (w/ CP) and default setting with first input stack as reference (w/o CP). (B-C) SSIM and NRMSE of reconstructed volumes with and without CP for large motion (B) and small motion (C). (*$P<0.001$, **$P<10^{-4}$) by paired t-test.

### 3.3 Performance of MA methods for random motion

We further simulated random motion that resembled real-world situations which was more challenging for MA. Among all 110 cases, CP showed the highest success rate of 100% in identifying the correct reference stack, compared to SVD-RSS and SVD-FS with success rates of 83.4% and 69.7%, respectively. As shown in Table. 2, there were obvious differences among the success rates along different axes. In SVD-FS, sagittal stacks had the highest success rate (motion in sagittal orientation was easier to detect), which was consistent with bias in RMI and BMI shown in Figure 3.

Finally, the reconstructed volumes using CP achieved the highest SSIM of 0.9840 and the lowest NRMSE of 0.1508, and fewer outliers compared to the SVD-FS and the default pipeline (Table 2). The exemplary reconstruction results in Figure 6 also indicated that the CP-based method achieved the best reconstruction quality with the highest similarity and the lowest error with respect to ground truth.

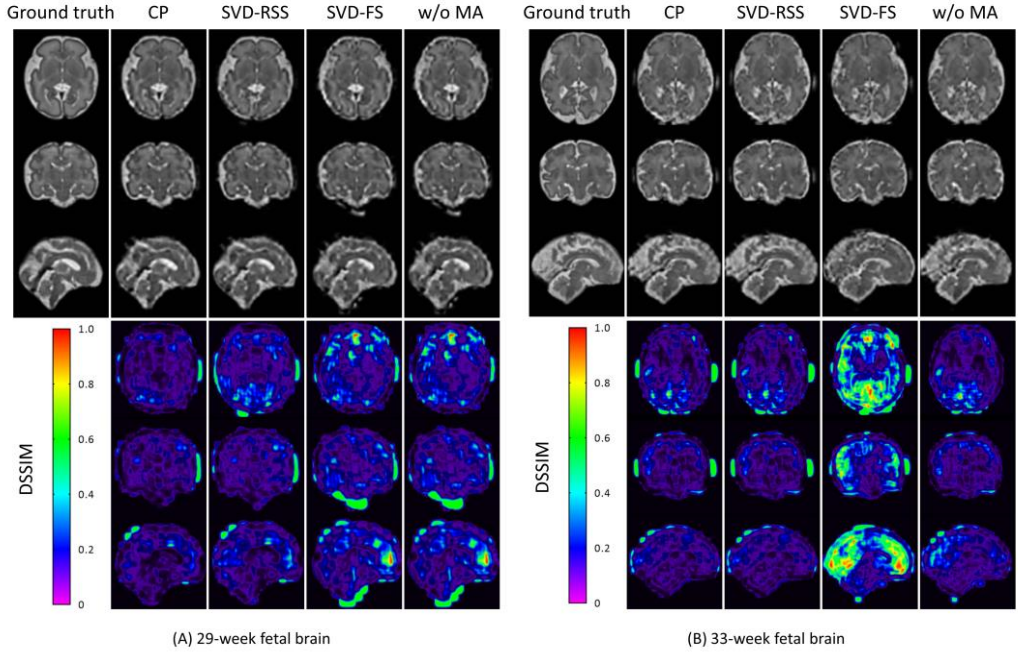

**Figure 6:** Comparison of reconstructed volumes and corresponding structural dissimilarity (DSSIM) maps of fetal brains at 31-week GA (A) and 36-week GA (B) using the selected minimum motion stack based on three MA methods and default pipeline (w/o MA).

**Table 2:** Comparison of the success rate and the quality of reconstructed volumes in terms of SSIM and NRMSE in the random motion experiment. Paired t-test was used to compare other methods to the CP method (*$P$<0.05, **$P$<0.01, ***$P$<0.001).

| MA method | Success rate | | | Reconstructed quality | |
|---|---|---|---|---|---|
| | Axial | Coronal | Sagittal | SSIM | NRMSE |
| **CP** | **100.00%** | **100.00%** | **100.00%** | **0.9840±0.0146** | **0.1508±0.0536** |
| SVD-RSS | 75.68% | 97.30% | 77.78% | 0.9836±0.0148* | 0.1528±0.0542 |
| SVD-FS | 86.49% | 27.03% | 97.22% | 0.9826±0.0155** | 0.1600±0.0579** |
| Without MA | - | - | - | 0.9816±0.0206*** | 0.1621±0.0702*** |

## 4. Discussion

Selecting the reference stack with the minimum motion is an important prior step in the retrospective volumetric reconstruction of fetal brain MRI. In this work, we proposed a CP decomposition-based method to determine the degree of motion using low-rank approximation and defined a MI. Experiments on the simulated linear and random motion showed the superior performance of the proposed method in comparison with two SVD-based methods, in terms of its sensitivity to motion and consistency of MIs among different orientations, as well as the success rate in determining the minimum motion stack and better quality of the reconstructed volume. The proposed method can be easily implemented as a prior step to any existing fetal brain MRI motion correction and super-resolution reconstruction pipeline given its high accuracy and low computational complexity.

The existing MA method utilized 2D SVD-based low-rank approximation that required factorizing the 3D stack into a 2D matrix by flattening all slices into vectors to be concatenated into a lengthy matrix with few rows but a huge number of columns [18]. This step could contaminate useful spatial information, preserving the structural and motion information along only one axis (the short axis), and thus lead to a less accurate assessment of fetal motion. More importantly, large baseline errors were introduced by the flattening of 3D stacks. For example, the flattened matrices of axial and coronal stacks were a collection of coronal and axial images, respectively (Supplementary Figure S5A-B), while the flattened matrix of sagittal stack contained rotated coronal images. (Supplementary Figure S5C). In the motion-free situation, these rotated coronal images were symmetric, and rows of the flattened sagittal matrix were linearly correlated, leading to a smaller baseline rank. When adding motion, the sagittal stack would have a greater increase of MI, which might explain the result that SVD-FS showed higher motion sensitivity in sagittal stacks, even outperformed CP (Figure 3-4). To address this limitation, we brought up a modified method (SVD-RSS) that interpolated a stack into an isotropic volume and then performed SVD on resliced images along three axes, to utilize full spatial information of the stack and reduce baseline bias among different stacks. Moreover, we found using the rank of data matrix to detect the similarity between slices was only suitable for simple structures, but it became difficult to assess motion for complex anatomical structures at larger GA. Therefore, we used the error between the original matrix and its low-rank component as MI. As shown in Figure 3 and Table. 1, the proposed SVD-RSS had better performance than conventional SVD-FS method as implemented in NiftyMIC. However, in the random motion experiment, the volumes reconstructed using SVD-RSS did not achieve desired quality, which might be related to the severe volume reduction caused by inter-slice motion. Also, the data matrix made by reslicing the interpolated volume along the non-principal axis would not be equivalent to that along the principal axis given the difference between in-plane and through-plane resolution, and therefore, SVD-RSS was only a pseudo-3D approach.

CP decomposition [27] was introduced to factorize a stack for motion assessment purposes for the first time. The proposed CP method shared the same core idea as SVD-RSS, both of which interpolated stacks into volumes to reduce structural bias among different orientations and then calculated the low-rank component to obtain motion information, yet CP provided a way for direct 3D factorization in a computationally efficient manner. The results of linear and random motion experiments demonstrated the advantages of treating the stack as a whole using CP decomposition, which gave higher motion sensitivity, success rate, and reconstruction quality compared to the two SVD-based methods. Moreover, CP also achieved lower computational time compared to SVD-FS. For instance, in a case of a fetal brain at 31 weeks of GA, the assessment of three stacks using SVD-FS cost 12 seconds, while the interpolation and factorization using CP only took about 6 seconds, and the computational time could be further shortened by reducing the number of rank-one tensors if needed.

There are several limitations in this work. First of all, since it is difficult to obtain the motion trajectory of fetal brain in real world [41], we used simulated linear and random motions in this study to test the performance of the motion estimation methods. Additionally, CP and SVD-RSS used the difference between original data and the low-rank component rather than traditional rank as the MI, other image attributes, such as contrast and signal-to-noise ratio, will also influence the evaluation metrics. This problem may be mitigated by performing SVD/CP on the relevant features, e.g., brain contours or high-frequency components of the image rather than the original image. Lastly, interpolation during volume-to-volume registration and other operations will blur images and reduce the sensitivity of motion detection [42]. Therefore, it would be more reasonable to consider image smoothness to penalize the drop in MI.

## 5. Conclusion

In this work, we proposed a CP-based method to assess motion and determine reference stack with the minimum motion for initialization of the fetal brain reconstruction pipeline. The proposed CP method utilized the difference between original data and its 3D low-rank approximation as MI, and showed superior performance compared to the previously used SVD-based method and its variations, in terms of sensitivity to motion, success rate, and baseline bias. This motion assessment method can serve as a simple and flexible plug-in to any registration-based motion correction algorithms to improve the volumetric reconstruction quality.

**Acknowledgment**

The work is supported by Ministry of Science and Technology of the People's Republic of China (2018YFE0114600, 2021ZD0200202), National Natural Science Foundation of China (81971606, 82122032), and Science and Technology Department of Zhejiang Province (202006140, 2022C03057).

**Supplementary Figure Captions:**

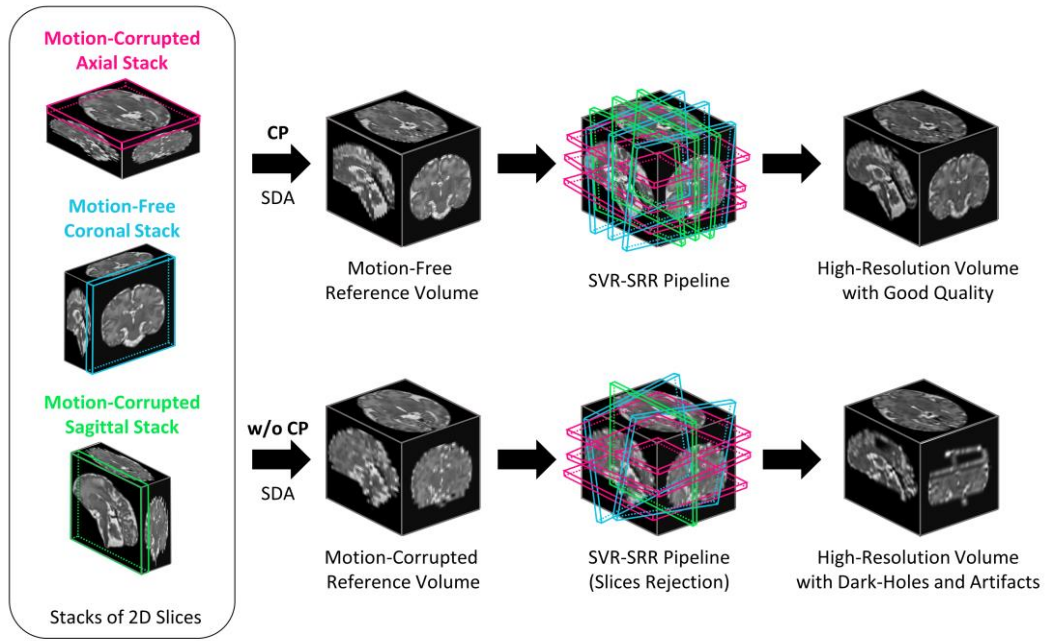

**Supplementary Figure S1:** Reconstruction pipelines using motion-free stack (coronal stack) as reference (top row) versus that using motion-corrupted stack (axial stack) as the reference (bottom row) in a 33-week fetus. The SVR-SRR pipeline with a motion-corrupted reference often fails due to slices rejection, resulting in voids and artefacts in the reconstructed volume.

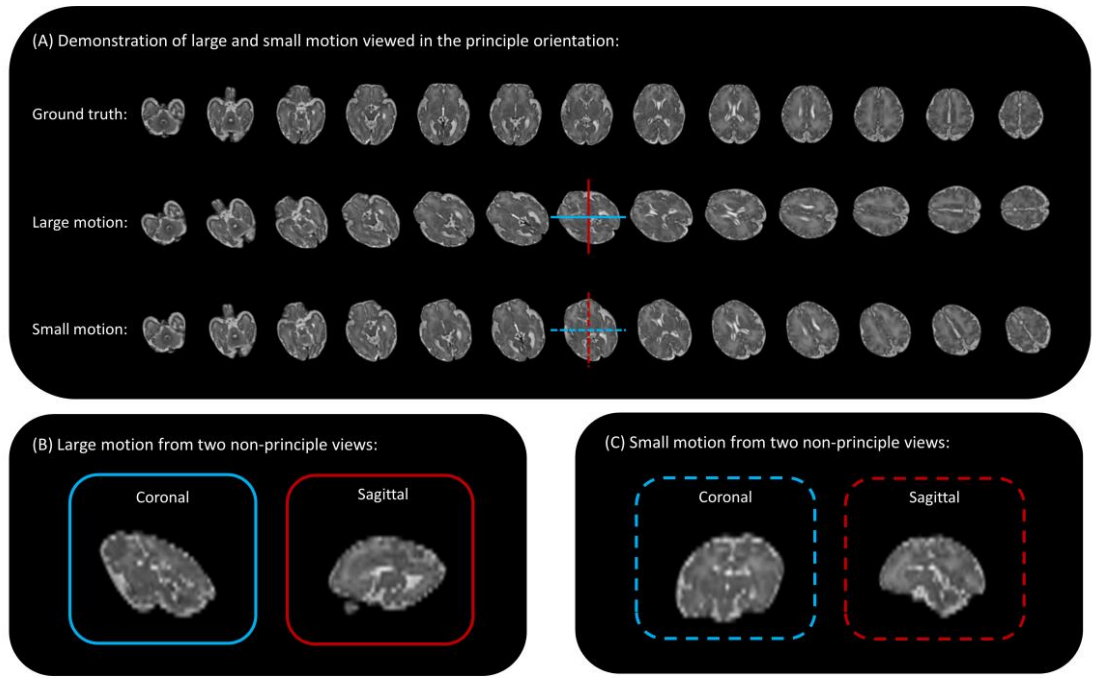

**Supplementary Figure S2**: Simulation of large and small motions in an axial stack of a 35-week-old fetal brain. (A) Motion-corrupted slices of large (middle row) and small motion (bottom row) in comparison with the motion-free slices (top row). (B-C) Coronal and sagittal views of the motion-corrupted fetal brain, corresponding to (A).

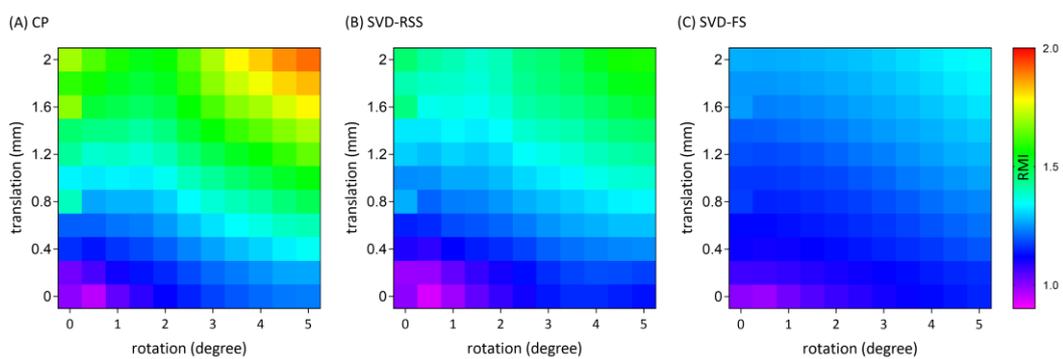

**Supplementary Figure S3:** Heatmaps of averaged RMI over three orientations at different combinations of motion parameters obtained using (A) CP, (B) SVD-RSS and (C) SVD-FS.

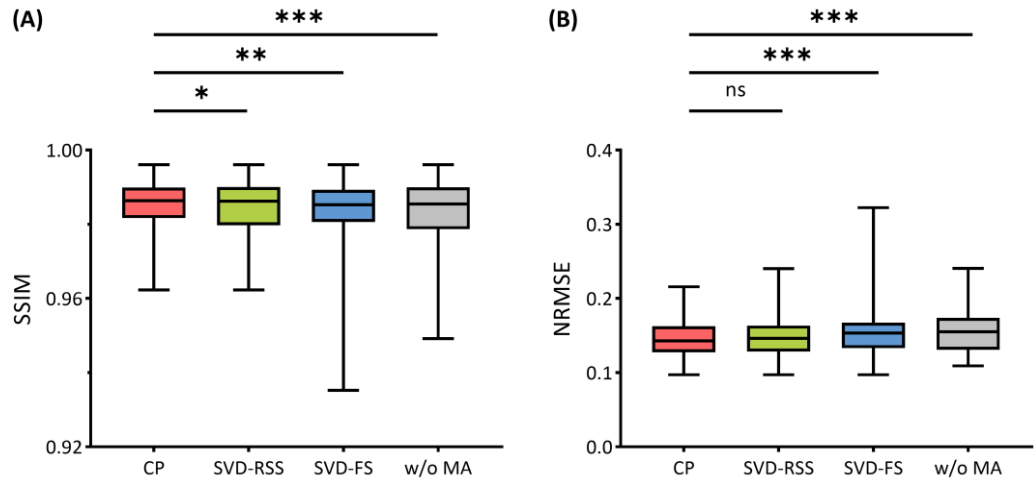

**Supplementary Figure S4:** Comparison of motion assessment methods in terms of the structural similarity (SSIM) and normalized root mean square error (NRMSE) of the reconstructed fetal brain volume using random motion-corrupted stacks respect to ground truth. (*$P<0.05$, **$P<0.01$, ***$P<0.001$) by paired t-test.

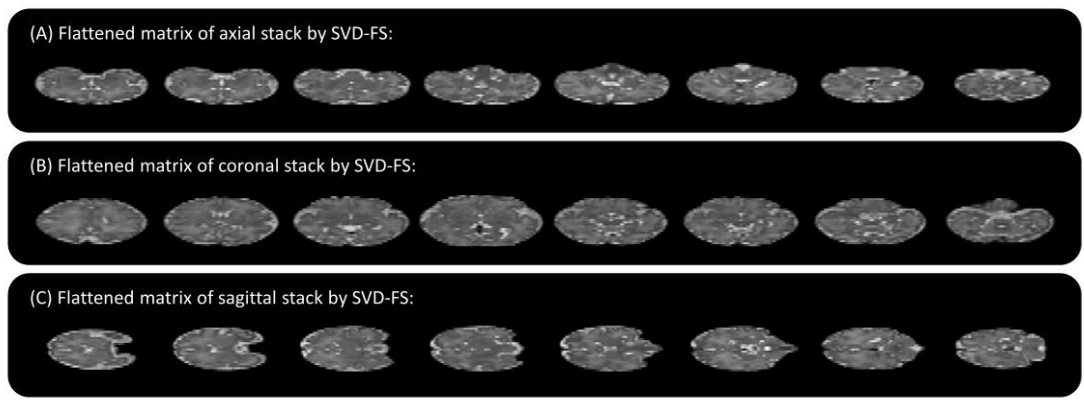

**Supplementary Figure S5:** 2D flattened matrices with few rows (n=24) but huge number of columns (in-plane image size=192*192=36864) obtained from the first step of SVD-FS for the (A) axial, (B) coronal and (C) sagittal stacks.

**Supplementary Table. S1.** Comparison of different motion assessment methods in terms of their RMI for large and small motions in three orientations.

| Motion assessment method | RMI of large motion | | | RMI of small motion | | |
|---|---|---|---|---|---|---|
| | axial | coronal | sagittal | axial | coronal | sagittal |
| CP | 1.516±0.181 | 1.592±0.213 | 1.509±0.182 | 1.268±0.117 | 1.224±0.105 | 1.180±0.081 |
| SVD-RSS | 1.235±0.054 | 1.354±0.075 | 1.248±0.064 | 1.113±0.066 | 1.146±0.059 | 1.096±0.054 |
| SVD-FS | 1.123±0.057 | 1.205±0.077 | 1.486±0.221 | 1.011±0.036 | 1.064±0.054 | 1.331±0.165 |

**Supplementary Table S2**: Comparison of different motion assessment methods in terms of their BMI in the motion-free stacks.

| Motion assessment method | BMI | |
|---|---|---|
| | axi-to-cor | axi-to-sag |
| CP | 0.943±0.034 | 0.991±0.035 |
| SVD-RSS | 0.982±0.050 | 0.995±0.030 |
| SVD-FS | 0.941±0.064 | 1.252±0.208 |